\documentclass[%
reprint,
onecolumn,
preprint,
 amsmath,amssymb,
 aps,
 prd,
floatfix,
]{revtex4-2}

\usepackage{graphicx}
\usepackage{dcolumn}
\usepackage{bm}
\usepackage{parskip}
\usepackage{xcolor}
\usepackage{url}
\usepackage{float}

\usepackage{hyperref}

\begin{document}
\preprint{GAME/23-QED-v3}

\title{Production of bound states of magnetic monopoles in high energy collisions at LHC}

\author{João Vitor Bulhões da Silva}
\email{joaovitor1729@gmail.com}
\author{Werner Krambeck Sauter}%
 \email{werner.sauter@ufpel.edu.br}
\affiliation{%
 Grupo de Altas e Médias Energias, Departamento de Física, Instituto de Física e Matemática,
Universidade Federal de Pelotas, Caixa Postal 354, 96001-970, Rio Grande do Sul, Brasil\\
}%
\date{\today}

\begin{abstract}
In this work, we present the studies carried out for the production of the monopolium at the LHC in ultraperipheral collisions for the processes $pp$ and $PbPb$. The monopolium is the bound state of a monopole-antimonopole pair. The study of the magnetic monopole in this characteristic state is justified because the coupling constant is very large, which allows us to suggest that this exotic particle can be produced in the bound state. The monopolium is characterized by a wave function arising from the numerical solution of the Schr\"{o}dinger equation with a modified Cornell potential. The monopolium is produced by a photon fusion production mechanism, with the Weizs\"{a}cker-Williams and Drees-Zeppenfeld expressions to describe the lead and proton equivalent photon distributions. We estimate a high production rate of monopolium production for $pp$ collisions with $\sqrt{s}=14$ TeV and $PbPb$ collisions with $\sqrt{s}=5.5$ TeV in LHC.
\end{abstract}


\maketitle
\section{\label{sec:level1}Introduction}

Magnetic monopoles have been the subject of great curiosity since they emerged from electromagnetic theory and have motivated the advancement of numerous experimental researches in the quest for this particle \cite{mavromatos2020magnetic}. Maxwell's equations have an electric-magnetic symmetry that is not presented without the discovery of magnetic charges. Another better motivation was given by Dirac, who showed that the existence of a single monopole is sufficient to explain the electrical charge quantization \cite{dirac:1931}. Therefore, if magnetic monopoles exist, the electric charge would be quantized, i.e., all the electric charges would be an integer multiple of a fundamental unit. The expression obtained by Dirac is the so-called Dirac Quantization Condition (DQC), which we can express as (in natural units):
\begin{equation}
    g=\frac{n}{2e}=n\left(\frac{e}{2\alpha}\right)=(68.5e)n,\quad n \in Z
    \label{eq.1.1}
\end{equation}
where $e$ is electric charge, $g$ is magnetic charge, $\alpha\equiv e^2\simeq1/137$ and $n$ is integer.

After some time of Dirac's contribution, Polyakov \cite{polyakov:1974} and 't Hooft \cite{tHooft:1974kcl} showed that magnetic monopoles arise as a solution of the field equation in Grand Unification Theory (GUT) \cite {PhysRevLett.43.1365}. In this theory, in the high energy regime, the electroweak and strong forces adopt a unified behavior, in a single symmetry group, and the generation of magnetic monopoles arises in the spontaneous symmetry breaking of this unification. Furthermore, 't Hooft and Polyakov showed that regardless of grand unification, the theory of particle physics must contain magnetic monopoles. Also an electroweak monopole is possible (see a review in\cite{cho2019physical}).

Depending on the model used for the GUT, the monopole mass prevision can fluctuate between $4\times 10^4$ GeV and astonishing $10^{17}$ GeV \cite{niessen:2001}. This be related to the idea that monopoles may have formed in the early Universe when the energy distribution was denser \cite{preskill:1984gd}. Thus, if a GUT were achievable in the early Universe after an inflationary period would remain a population of the magnetic monopoles as relics \cite{Kibble:1980mv}. Consequently, due to the decays of magnetic monopoles produced in the Big Bang, it is possible to observe them in the Cosmic Microwave Background (CMB) \cite{Eto:2020hjb}. A possible explanation for the absence of experimental evidence for monopoles was considering that they cannot be freely detected, given the fact of the strong magnetic coupling constant, giving us the possibility that they can be produced in a bound state called monopolium \cite{Epele:2007ic, Barrie:2016wxf_2, barrie2021searching}.

From an experimental point of view, there are several attempts to measure the physical properties of magnetic monopoles~\cite{Patrizii_2015}. One of them, the Monopole and Exotics Detector Experiment at the LHC (MoEDAL) is an experiment dedicated to the search for monopoles, ions, and other highly ionizing particles in high-energy collisions, located at Point 8 of the LHC ring \cite{doi:10.1142/S0217751X14300506}. Currently, no experiment has been able to generate signals of the production of this particle \cite{PhysRevLett.123.021802}, giving us only limits of mass and charge \cite{PhysRevLett.118.061801}. Furthermore, the search of GUT magnetic monopoles in current particle accelerators is excluded due to high mass, which suppresses the cross-section.

We consider the central exclusive production process via photon fusion, producing a monopole and an antimonopole in a bound state, the monopolium, which is described by a Cornell-like potential. The article is organized as follows. In the next section \ref{monopolium}, we present an overview of the theory of magnetic monopoles with a short review of the cross-section production of the monopolium. In section \ref{ultraperipheral}, we present the mechanism of central production in ultraperipheral collisions with the central system of particles created by a pair of high-energy photons. The results of the cross-sections are shown and discussed in section \ref{results}. Finally, in section \ref{summary}, a summary and the conclusions are presented.

\section{Monopolium production and decay}\label{monopolium}

The magnitude of the force of attraction between two monopoles with opposite magnetic charges, according to the DQC, is $(68.5)^2\approx4{,}692.25$ times greater than the force of attraction between an electron-positron pair. For this reason, we are looking for magnetic monopoles in their bound state. Some previous works assume that this particle can be more easily detected in its bound state than freely \cite{Epele:2007ic, epele2009MonprophofusLarHadCol}. Thus, if we consider a monopole-antimonopole pair, due to the value of the coupling constant, it is possible to have the monopolium production. Subsequently, there will be the monopolium decay into two photons.

Therefore, we revisit the production of monopolium via the photon fusion mechanism and its subsequent two-photon decay. In particular, we will show the bound states production of magnetic monopoles via photon fusion in high-energy collisions of $pp$ and $PbPb$, with energy available at the LHC. In this work, the monopolium is an unstable intermediate state (a resonance) with a small decay width and the cross-section for this process will have a peak in the monopolium mass \cite{PhysRev.49.519}. Previous studies of monopolium production and magnetic monopole pairs on colliders are found in \cite{jean,bruna,baines2018monopole,dougall2009dirac,Epele:2007ic,epele2009MonprophofusLarHadCol,PhysRevD.60.075016,PhysRevD.57.R6599}.

The resonance cross-section formula for the monopolium production can be written by:
\begin{equation}
    \sigma(\gamma\gamma\rightarrow M)=\frac{4\pi}{\hat{s}}\frac{M^2\Gamma(\sqrt{\hat{s}})\Gamma_M}{(\hat{s}-M^2)^2+M^2\Gamma^2_M},
    \label{M}
\end{equation}
where $\hat{s}$ is the center of mass energy, the width decay of monopolium is $\Gamma_M=10\text{ GeV}$ \cite{Epele:2007ic}, $M= 2m+E_{bind}$ is the monopolium mass which depends only on the masses of the constituent monopoles and the binding energy $E_{bind}$, given by the solution of the radial Schr\"{o}dinger equation for the potential $V(r)$, which describes the bound state of the magnetic monopoles.

\begin{figure}[h]
    \centering    \includegraphics{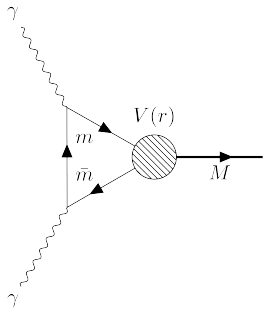}
    \caption{Production of monopolium by photon fusion. In the diagram, we have that $M$ represents the monopolium and $V(r)$ is the potential energy between the monopole ($m$) and the antimonopole ($\bar{m}$).}
    \label{2f1M}
\end{figure}

The production width $\Gamma(\sqrt{\hat{s}})$ is
\begin{equation}    \Gamma(\sqrt{\hat{s}})=\frac{32\pi\alpha_{mag}^2}{M^2}|\psi_M(0)|^2,
    \label{pwidth}
\end{equation}
where $\alpha_{mag}$ is the magnetic coupling constant and $\psi_{M}(0)$ is the wave function in the origin of the bound state of the monopole-antimonopole pair, see Fig. \ref{2f1M}. In particular, two formulations for the magnetic coupling constant will be considered: the first is given by Eq. (\ref{eq.1.1}) that is equal to $\alpha_{mag}=g^2/4\pi$ and the another one is $\alpha_{mag}=(\beta g)^2/4\pi$, which $\beta=(1-M^2/\hat{s})^{1/2}$ is the velocity of the monopole \cite{rajantie2012introduction}.

\begin{figure}[h]
    \centering
    \includegraphics{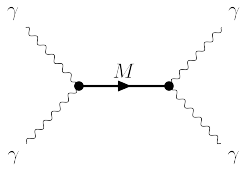}
    \caption{Diagrammatic description of the monopolium production ($M$) and its subsequent decay into $\gamma\gamma$. }
    \label{fffM}
\end{figure}

The monopolium can decay into two (or more) photons shortly after the bound state has been formed. The Feynman diagram of this process can be seen in the Fig. \ref{fffM}.

The cross-section of the above decay is similar to the monopolium production process \cite{Epele2012}. We have,
\begin{equation}
    \sigma(\gamma\gamma\rightarrow M\rightarrow \gamma\gamma)=\frac{4\pi}{\hat{s}}\frac{M^2\Gamma^2(\sqrt{\hat{s}})}{(\hat{s}-M^2)^2+M^2\Gamma^2_M},
    \label{2ff}
\end{equation}
where $\Gamma(\sqrt{\hat{s}})$ is the production width of the monopolium decaying into two photons.

The bound state is characterized by a wave function, solution of Schr\"{o}dinger equation with an interaction potential. A model for this potential is presented in \cite{Barrie:2016wxf_2}, where Zwanziger's dual electromagnetic formulation and lattice gauge theory were considered, and, due to the high value of the coupling constant, this approach yields a linear term in the potential justifying the dynamical behavior of a system with large coupling constant. Added with this potential, there is the usual weak Coulomb potential. A similar model used for monopolium search \cite{azevedo2022there} uses a procedure in which the mesons are described as two heavy quarks carrying opposite magnetic charges within a confining state and, finally, the effective potential is also presented as a linear plus Coulombian contribution. In a recent work \cite{abreu2020effective}, based on Nambu model \cite{nambu1974strings} for heavy-effective field theory, whose effective Lagrangian that arises from studies of heavy-meson has been adapted to monopolium as two- and four-body bound states. The result of this combination is high-precision QCD calculations, where the potential between a quark-antiquark pair is well approximated by the Cornell potential \cite{Bali:1996bw}.

\begin{figure}[h]
    \centering    
    \includegraphics[width=\columnwidth]{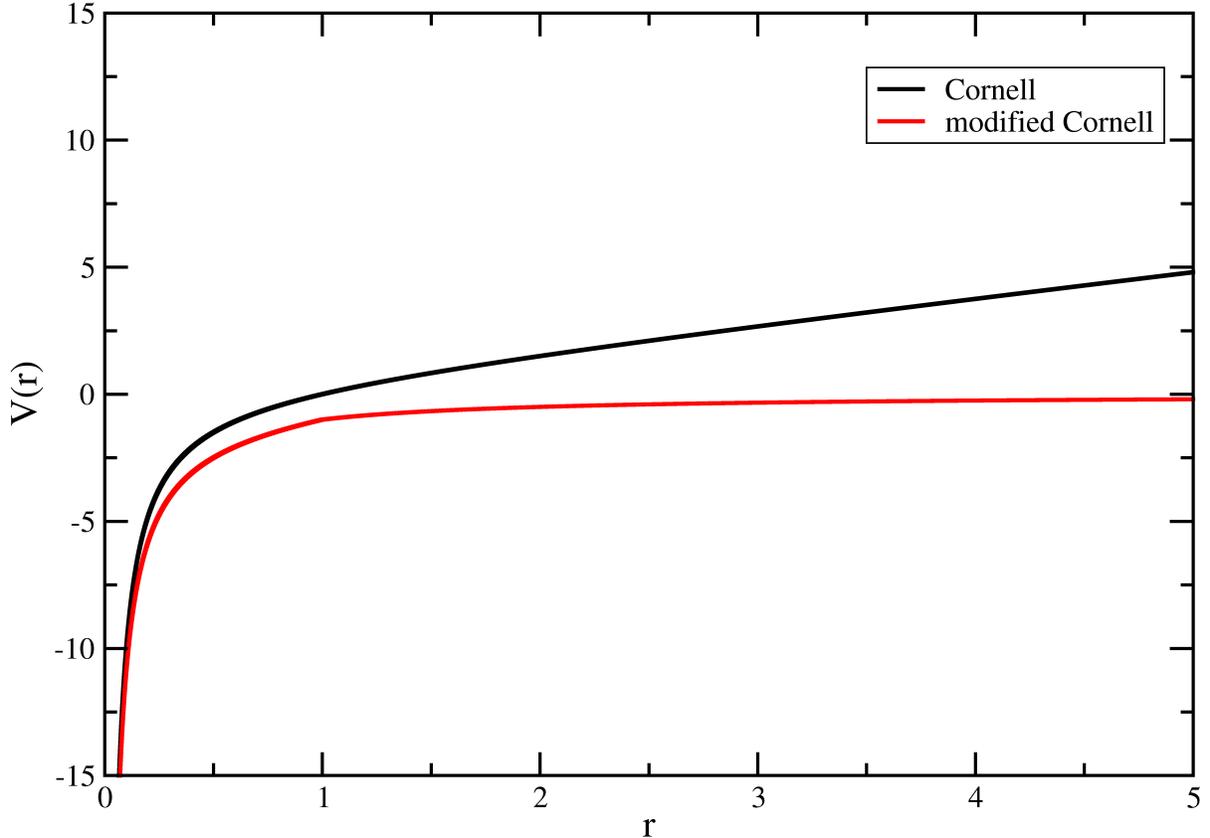}
    \caption{Behavior of the Cornell potential and modified Cornell potential $V(r)$ that describes the bound state of the monopole-antimonopole pair.}
    \label{pot}
\end{figure}

These results motivate us to assume that a pair of point-like magnetic monopoles can have the interaction described by the Cornell-like potential,
\begin{equation}
    V(r)=-\frac{g^2}{4\pi r}+\frac{\ln(2g^2)}{a^2}r,
\end{equation}
where $g$ is the magnetic charge and $a=1/0.52 \text{ GeV}^{-1}$ is a lattice parameter. We can note on the Fig. \ref{pot} that the potential energy grow up with the distance $r$ because of the linear part of the potential. This behavior describes a confinement regime because the potential $V(r)$ also acts for $r\rightarrow \infty$. For monopoles to exist in a situation in which the linear term of the potential is limited, and for this particle to be found in the unconfined state, we need the potential:
\begin{equation} \label{potcornmod}
 V(r)= \begin{cases}
        -\kappa(r_0-r)-g^2/4\pi r,&  r\leq r_0 \\
        -g^2/4\pi r,&  r >r_0
       \end{cases}
\end{equation}
where $\kappa=\ln(2g^2)/a^2$, also known as string tension. Thus, the Cornell potential modified (\ref{potcornmod}) ensure that 
\begin{equation}
    \lim_{r\rightarrow \infty} V(r)=0.
\end{equation}

Moreover, we can specify the $r_0$ value that limits the contribution of the linear term of the potential. For this purpose, it is sufficient to assert that $V(r_0)=0$, 
\begin{equation}
V(r_0)=-\frac{g^2}{4\pi r_0}+\frac{\ln(2g^2)}{a^2}r_0=0,
\end{equation}
giving us,
\begin{equation*}
  r_0=g\sqrt{\frac{1}{4\pi \kappa}}.  
\end{equation*}

Therefore, the wave function $\psi_M(0)$, used in the Equation (\ref{pwidth}), will be calculated using the modified potential of Equation (\ref{potcornmod}). It is known that there is no analytical solution for the Schr\"{o}dinger equation for the potential in question. However, it is possible to perform this calculation using numerical methods, such as the Numerov one (see \cite{giannozzi2013numerical} for a review). We use this method to solve the radial Schr\"{o}dinger equation for the modified Cornell potential, which allows us to obtain the value of wave function at the origin as $\psi_{M}(0)=1,803.85 \text{ a.u}$ and the binding energy of the ground state as $E_{bind}=8,735 \text{ Ry}$.

\section{Ultraperipheral hadron collisions}\label{ultraperipheral}

Ultraperipheral collisions are a process in which two fast-moving charged hadron projectiles ($h_A$ and $h_B$) are separated by a large impact parameter $b$ that should be $b>R_{h_A}+R_{h_B}$, where $R_{h_i}$ is the hadron radius. In the ultra-relativistic limit, the strong electromagnetic field generated can be replaced by an equivalent flux of photons\cite{terazawa1973Twophoproparprohig,Bertulani_2015,contreras2015ultra} and its intensity is directly related to the number of equivalent photons that involves the projectiles, which is proportional to the square of atomic number $Z^2$ \cite{NYSTRAND2005470, vonweizsacker1934Rademicolverfasele, williams1934Nathigparpenradstaionradfor}. We will use the Weizs\"{a}cker-Williams (WW) and Dress-Zeppenfeld (DZ) expressions (see below).

The Weizs\"{a}cker-Williams photon number distribution\cite{vonweizsacker1934Rademicolverfasele, williams1934Nathigparpenradstaionradfor} is given by 
\begin{equation}
f(x)=\frac{Z^2\alpha_{el}}{\pi}\frac{1}{x} \left[2Y K_0(Y)K_1(Y)-Y^2(K_1^2(Y)-K_0^2(Y))\right],
\label{eq. 9}    
\end{equation}
where $K_0$, $K_1$ are modified Bessel functions, $Y=xM_Ab_{min}$, $x$ is the energy fraction of the photon, $M_A$ is the hadron mass of the beam and $b_{min}$ is the minimum impact parameter. In the proton case, $b_{min}=0.7\text{ fm}$ and in the ion case, $b_{min}=14.2 \text{ fm}$.

Furthermore, another proton equivalent photon flux was computed by Dress and Zeppenfeld \cite{drees1989ProSupParElaepCol}, where they take into account the form factor of the electric dipole,  
\begin{equation}
f(x)=\frac{\alpha_{el}}{2\pi x}[1+(1-x)^2] \left[\ln A-\frac{11}{6}+\frac{3}{A}-\frac{3}{2A^2}+\frac{1}{3A^3}\right],
\label{eq. 10}
\end{equation}
where
\begin{equation}
    A=1+\frac{0.71 \text{GeV}^2}{Q^2_{min}},\  Q_{min}^2=\frac{m_p^2x^2}{1-x}.
\end{equation}

\begin{figure}[h!]
    \centering
    \includegraphics[width=0.8\columnwidth]{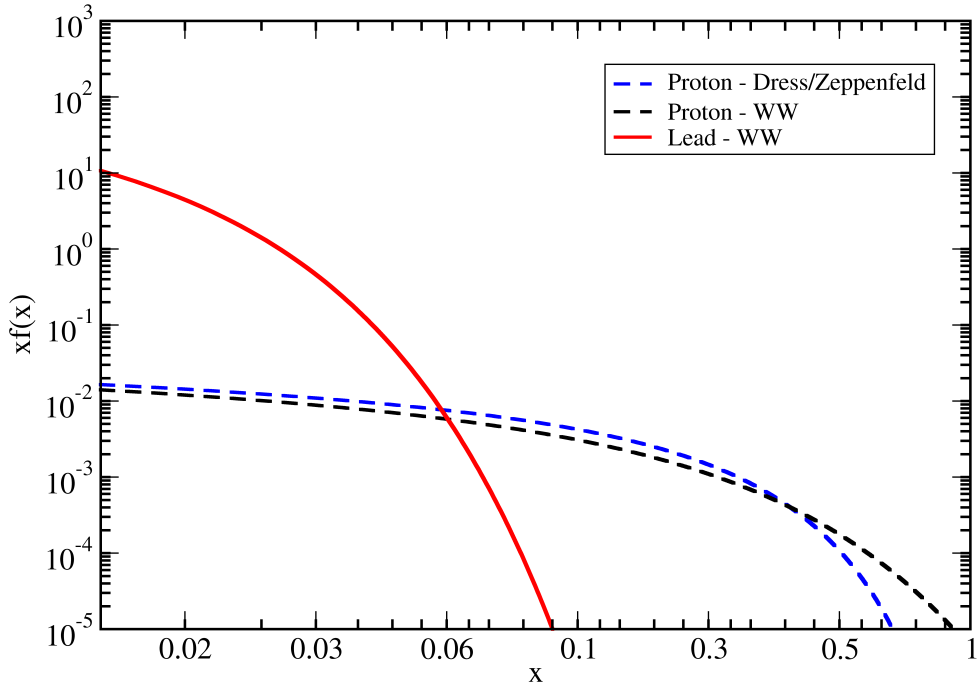}
    \caption{Number of equivalent photons of proton and lead as a function of the fraction of energy carried by photons.}
    \label{nfoton}
\end{figure}

In Fig.\ref{nfoton}, we can see that the $Z^2$ factor in Eq. (\ref{eq. 9}) is very important to understand the behavior of the equivalent photon spectrum from different projectiles. In the case of lead, the number of equivalent photons is greater than that of the proton, with a difference of almost three orders of magnitude, but the fraction of energy carried by photons is small when compared with the proton. The disadvantage of using heavy ions is the very low energy of the photons, which can be a problem if we are interested in a massive resonance with high energy.

We will consider a coherent emission of photons, leaving the projectiles intact in the final state. Photons can interact in several ways, but we will only consider the case where photons will interact with each other through photon fusion $\gamma\gamma$, generating a massive resonance $M$.

We also consider the elastic contribution for the cross-section, since the interaction is purely coherent. The total cross-section for two-photon interaction can be written as \cite{NYSTRAND2005470}:
\begin{equation} \label{eq cs_el}
\sigma_{tot}=\int_{M^2/{s}}^1 dx_1\int^1_{M^2/{s}x_1}dx_2 f(x_1)f(x_2) \sigma_{\gamma\gamma\rightarrow X}(x_1 x_2{s}),
\end{equation}
where $M$ is the mass of the central produced system, $s$ is the center of mass energy of the projectiles, $x_{i}$ is the fraction of energy carried by the photon, $f(x_i)$ is the equivalent photon spectrum produced by a charged particle and $\sigma_{\gamma\gamma\rightarrow X}$ is two-photon cross-section of production of a $X$ state, Eqs. (\ref{M}) and (\ref{2ff}).

\section{Results}\label{results}

We calculate the total cross-section of monopolium and its decay into $\gamma\gamma$. The total cross-section is a function of the monopolium mass, so the energy of the center of mass remains fixed. We compare the processes of collisions at the LHC with center of mass energy of $\sqrt{s} =$ $14$ TeV for $pp$, $\sqrt{s} =$ $5.5$ TeV for $PbPb$ collisions with the integrated luminosity with values $150.0 \text{ fb}^{-1}$ and $5.0 \text{ nb}^{-1}$, respectively.

\begin{figure}[h!]
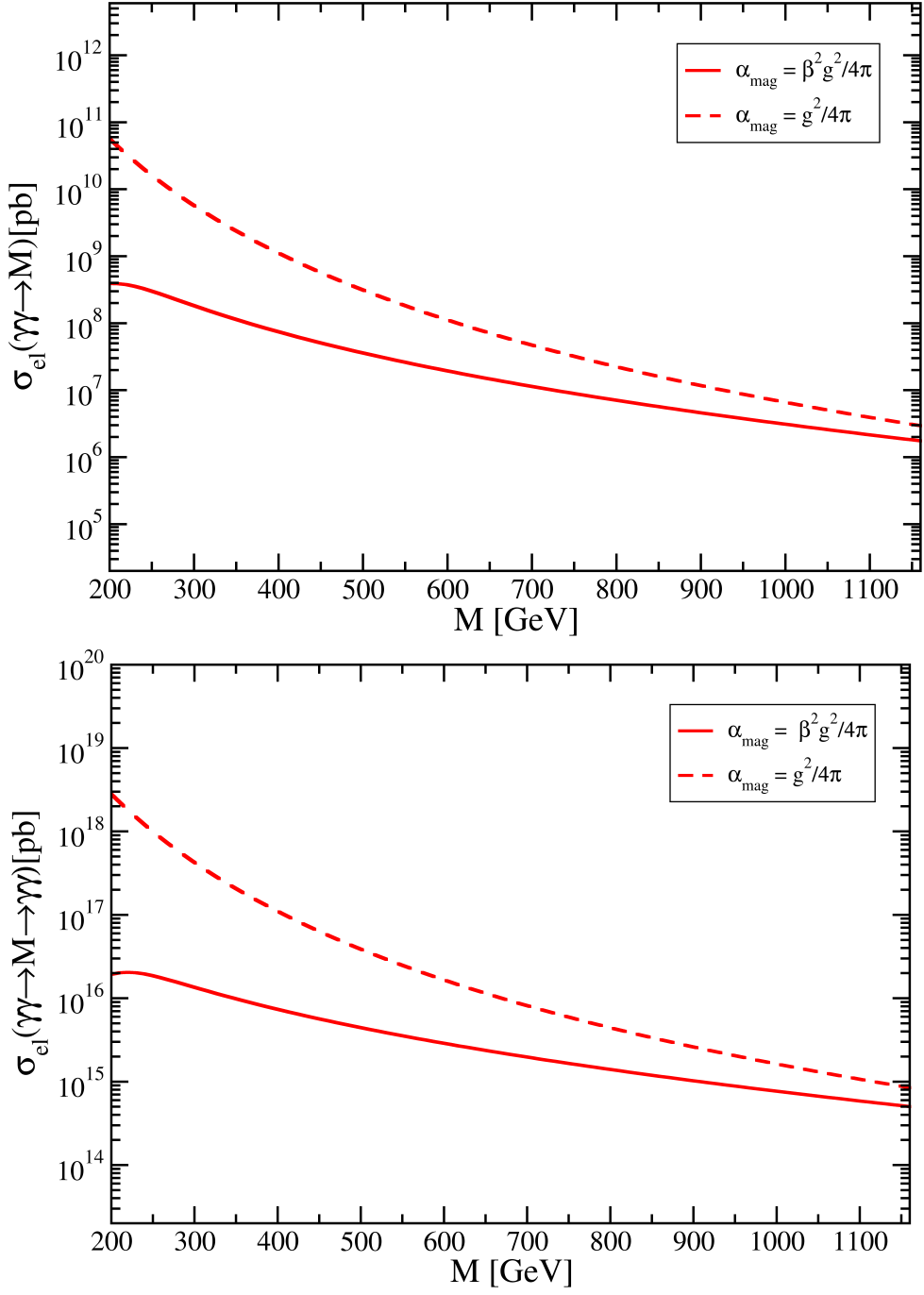

\centering
    \includegraphics[width=0.8\columnwidth]{ppww.pdf}\\
    \includegraphics[width=0.8\columnwidth]{ppwwgg.pdf}
\caption{Total cross-section of proton-proton collisions with $\sqrt{\hat{s}}=14$ TeV for the photoproduction of monopolium (up) and its decay into two photons (down) as a function of the monopolium mass $M$, where was considered the Weizs\"{a}cker-Williams photon distribution.}\label{label-a}
\end{figure}

\begin{figure}
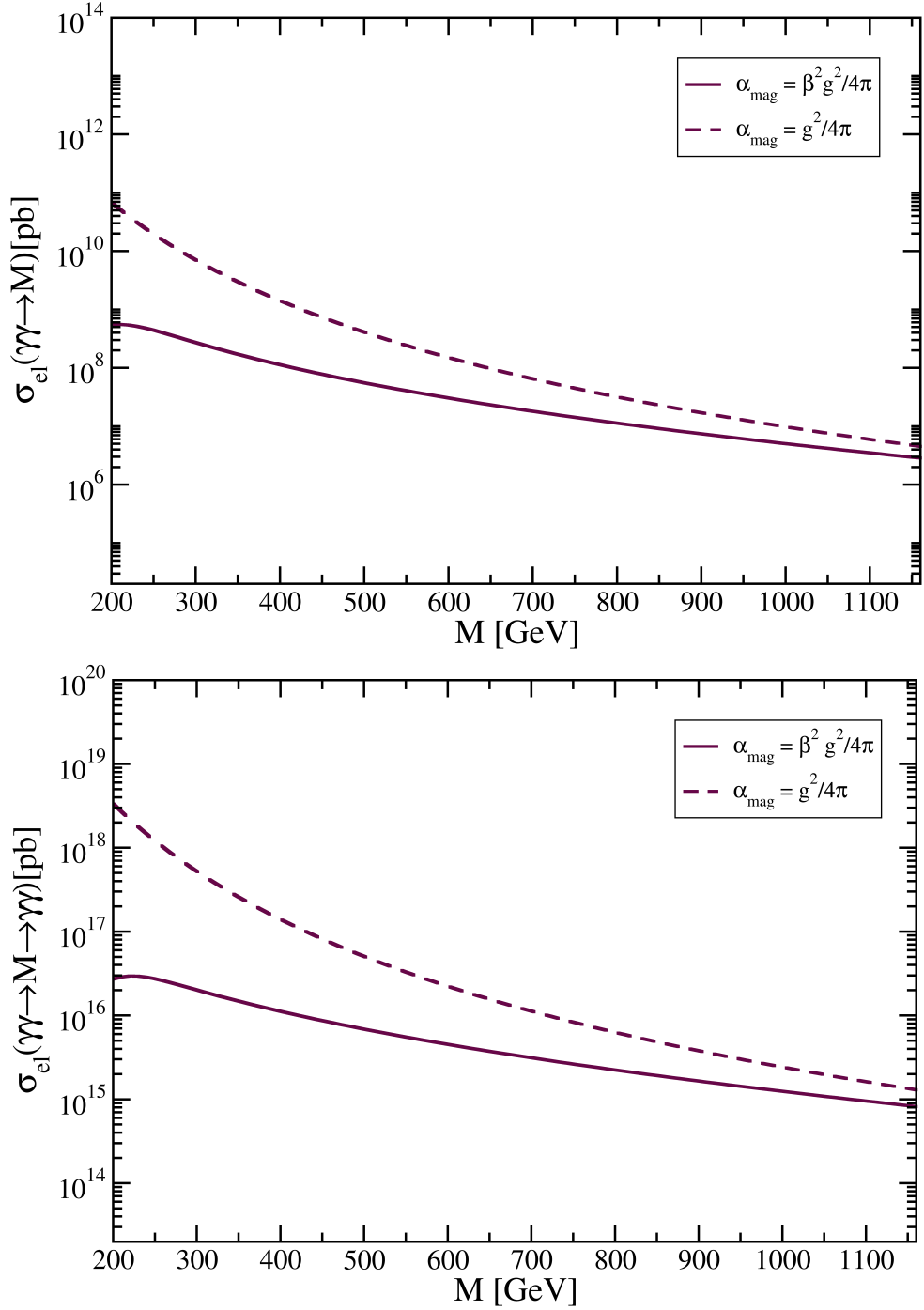

\centering
    \includegraphics[width=0.8\columnwidth]{ppDZ.pdf}\\
    \includegraphics[width=0.8\columnwidth]{ppDZgg.pdf}
\caption{Total cross-section of proton-proton collisions with $\sqrt{\hat{s}}=14$ TeV for the photoproduction of monopolium (up) and its decay into two photons (down) as a function of the monopolium mass $M$, where was considered the Dress-Zeppenfeld photon distribution.}\label{label-b}
\end{figure}

\begin{figure}
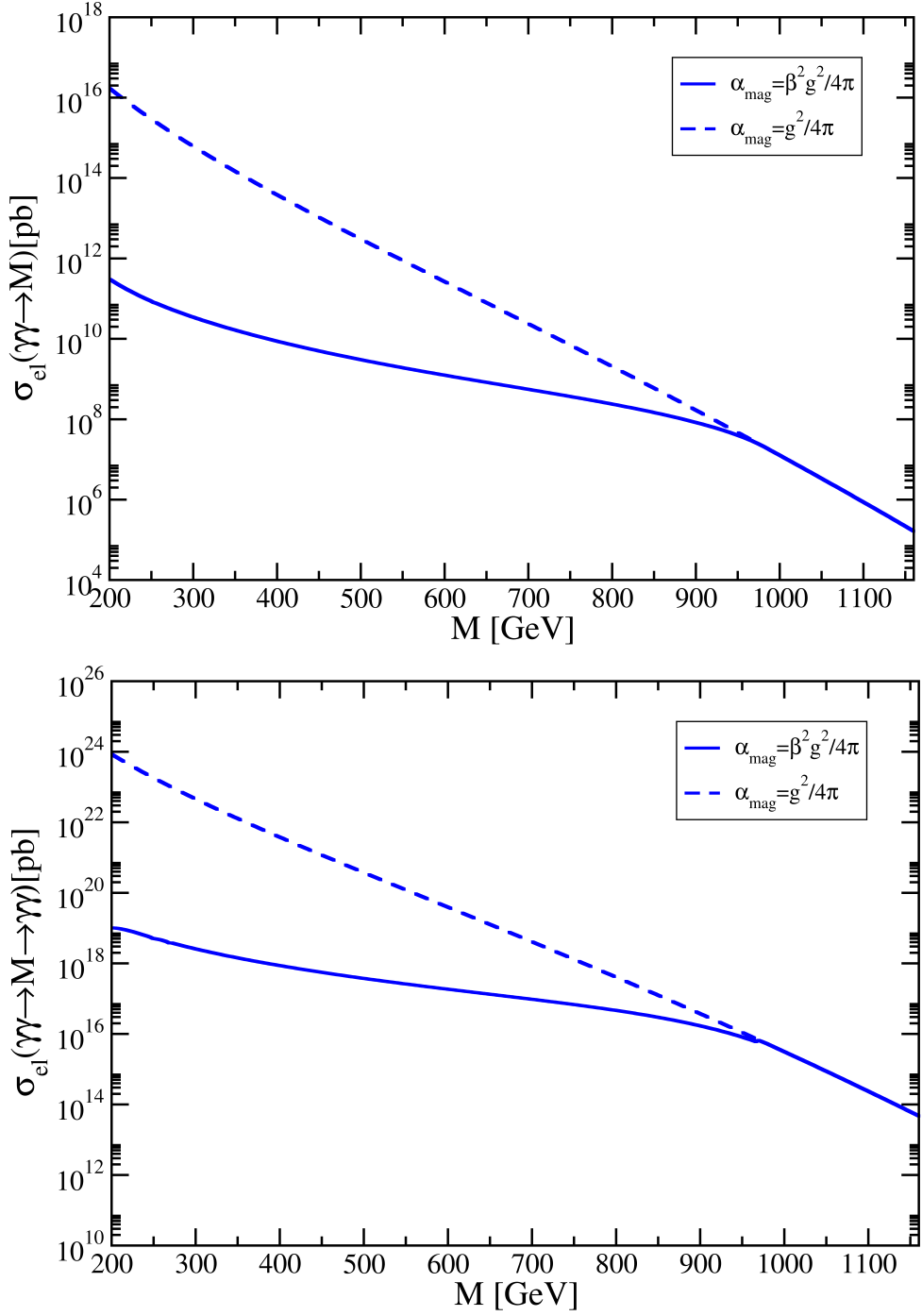

\centering
    \includegraphics[width=0.8\columnwidth]{PbPbggM.pdf}\\
    \includegraphics[width=0.8\columnwidth]{PbPbggMgg.pdf}
\caption{Total cross-section of lead-lead collisions with $\sqrt{\hat{s}}=5.5$ TeV for the photoproduction of monopolium (up) and its decay into two photons (down) as a function of the monopolium mass $M$, where was considered the Weizs\"{a}cker-Williams photon distribution.}\label{label-c}
\end{figure}

The monopole mass limits are $415 \text{ GeV}<m<2{,}420\text{ GeV}$ and, by using the binding energy $E_{bind}$ value, we can define the range of monopolium mass limits for this theory as $200 \text{ GeV}<M<1{,}160\text{ GeV}$. The calculation of the Eq. (\ref{eq cs_el}) gives us the total cross-section for the photoproduction process. In the following, we present the estimates of the production and decay of monopolium in $pp$ collisions: in Fig. (\ref{label-a}) was used the Weizs\"{a}cker-Williams photon flux and in the Fig. (\ref{label-b}) the Dress-Zeppenfeld photon distribution. Moreover, Fig. (\ref{label-c}) presents the total cross-section for monopolium production and decay in $PbPb$ collisions, with Weizs\"{a}cker-Williams photon flux.

We consider two expressions for the coupling constant: the first is that the intensity of the coupling constant depends on the monopole velocity $\alpha_{mag}=(\beta g)^2/4\pi$, and the second depends only on the square of the magnetic charge $\alpha_{mag}=g^2/4\pi$. The direct choice of coupling constant only affects the intensity of cross-sections. Furthermore, as we can see in Figs.  (\ref{label-a}), (\ref{label-b}) and (\ref{label-c}), the difference in cross-section for the two approaches to the coupling constants becomes small as the mass $M$ increases. For the case of $PbPb$ collisions, there is no difference in the cross-section at $M\simeq 950 \text{ GeV}$, implying that the different couplings make the same contribution in this mass limit.

The theoretical approach shows that monopolium mass can be smaller than the monopole mass, due to the bound state energy. A large part of the energy available in the process is used to bind the monopole-antimonopole pair, and the remaining energy will compose the monopolium mass. Furthermore, the photon that emerges from the decaying product of the monopolium has an energy equal to $M/2$.

According to the displayed results, we can observe that the processes involving collisions of the type $PbPb$ present a higher cross-section than the processes involving collisions of the $pp$. This behavior can be explained by observing Fig. (\ref{nfoton}), where it is possible to notice that lead has a large number of equivalent photons when compared to the proton, despite its integrated luminosity in the LHC being lower than when compared with the luminosity for collisions of the type $pp$ \cite{luminosidade, Bruce_2020}. Now, comparing the present results with the previous article\cite{jean}, where the Coulombian potential is employed, we have larger cross-sections in proton-proton collisions than in ion-ion collisions. The differences come from the distinct bound energies: the Coulomb potential has a smaller bound energy than the Cornell-like potential used in this work. The photon energy is used to producing the particle/antiparticle pair, but the bound energy is large, resulting in a light mass bound state. The distribution of photons for nuclear projectiles favors the production of light mass states, explaining the difference between the present work and the previous one\cite{jean} and the larger cross section for ion-ion collisions.

In tables \ref{tab1} and \ref{tab2} we show the production rate of the monopolium and monopolium as a two-photon resonant state, respectively, for the luminosity of hadron collisions in LHC energies. Due to the large cross-sections, the predicted number of events is also large.

\begin{table}[h!]
\caption{Event rates for the monopolium production.}
\centering
\begin{tabular}{ |c|c|c|c| }
 \hline
 \multicolumn{4}{|c|}{$\gamma\gamma\rightarrow$ Monopolium - $\alpha_{mag}=\beta^2 g^2$} \\
 \hline
  $M$ (GeV) & events $pp$ - WW & events $pp$ - DZ & events $PbPb$ - WW  \\
 \hline
 $1{,}160$ & $2{.}56\times 10^7$  & $4{.}26\times 10^7$&$6{.}83\times 10^2$\\
 \hline
 \multicolumn{4}{|c|}{$\gamma\gamma\rightarrow$ Monopolium - $\alpha_{mag}=g^2$}\\
 \hline
 $M$ (GeV) & events $pp$ - WW & events $pp$ - DZ & events $PbPb$ - WW  \\
 \hline
 $1{,}160$ & $4{.}41\times 10^7$   & $6{.}76\times 10^7$ &$6{.}83\times 10^2$\\
 \hline
\end{tabular}
\label{tab1}
\end{table}

\begin{table}[h!]
\caption{Event rates for the production of monopolium as a two-photon resonant state.}
\centering
\begin{tabular}{ |c|c|c|c| }
 \hline
 \multicolumn{4}{|c|}{Monopolium$\rightarrow\gamma\gamma $ - $\alpha_{mag}=\beta^2 g^2$} \\
 \hline
  $M$ (GeV) & events $pp$ - WW & events $pp$ - DZ & events $PbPb$ - WW  \\
 \hline
  $1{,}160$ & $7{.}51\times 10^{15}$  & $1{.}22\times 10^{16}$&$1{.}97\times 10^{11}$\\
 \hline
 \multicolumn{4}{|c|}{Monopolium$\rightarrow\gamma\gamma $ - $\alpha_{mag}=g^2$} \\
 \hline
 $M$ (GeV) & events $pp$ - WW & events $pp$ - DZ & events $PbPb$ - WW  \\
 \hline
 $1{,}160$& $1{.}26\times 10^{16}$ & $1{.}93\times 10^{16}$ & $1{.}97\times 10^{11}$\\
 \hline
 \end{tabular}

\label{tab2}
\end{table}

\section{Summary and conclusions}\label{summary}

The theoretical justification for the existence of monopoles is that they add symmetry to Maxwell's equations and explain the charge quantization. Dirac showed that the existence of monopoles in the Universe may explain the discrete nature of electric charge. Magnetic monopoles have been predicted by several theories, such as the Grand Unification Theory (GUT), Quantum Gravity Theory, and Electroweak Theory.

We consider the production of the monopolium in two ways: in the first process, the incident-charged hadrons emit photons that interact with each other, producing the monopolium as the final state; in the second one, the monopolium decays into two photons in the final state. To calculate the production rate of the monopolium, it is necessary to use the wave function $\psi_M(0)$ of the bound state. This wave function is calculated from a (more realistic) modified Cornell potential, instead of a Colombian potential with a hard core\cite{Epele2012}, used previously in the same observables\cite{jean}. We found a remarkably large cross-section (and predicted numerous events in LHC) for both final states, for both $pp$ and $PbPb$ collisions, in comparison with the previous results \cite{jean}. The result indicates that a more realistic treatment of the bound states of magnetic monopoles is necessary to constrain the search for this exotic particle in ultraperipheral collisions.

\begin{acknowledgments}
The authors thank the Grupo de Altas e Médias Energias for the support in all stages of this work. J.V.B.S. thanks to CAPES, CNPq, FAPERJ and INCT-FNA (process number 464898/2014-5) for the financial support during the development of this work.
\end{acknowledgments}

\bibliography{apssamp.bib}

\end{document}